\documentclass{cimento}

\usepackage{graphicx}
\title{Coulomb law and energy levels
in a superstrong magnetic field}
\author{M.I.~Vysotsky\from{ins:x}}
\instlist{\inst{ins:x} Institute of Theoretical and Experimental
Physics, Moscow, Russia}

\PACSes{\PACSit{10.11} \PACSit{30.31}}

\begin{document}

\maketitle

\begin{abstract}
Analytical expression for the Coulomb potential in the presence of
superstrong magnetic field is derived. Structure of hydrogen
levels originating from LLL is analyzed.
\end{abstract}

\section{Introduction}

The long awaited discovery of Higgs boson is planned during the
next two years at LHC. For the first time what is called now the
Higgs phenomenon was used in the Ginzburg--Landau phenomenological
theory of superconductivity to expel the magnetic field from a
superconductor.

Quite unexpectedly in the superstrong magnetic field a photon also
gets (quasi) mass. In this talk we have discussed this phenomenon
and how it effects the atomic energy levels. The talk is based on
papers \cite{1}.

In what follows the strong magnetic field is $B>m_e^2 e^3$; the
superstrong magnetic field is $B>m_e^2/e^3$; the critical magnetic
field is $B_{cr}=m_e^2/e$ and we use gauss units: $e^2 = \alpha =
1/137$.

Landau radius of an electron orbit in the magnetic field $B$ is
$a_H=1/\sqrt{eB}$ and it is much smaller than the Bohr atomic
radius for $B>>e^3 m_e^2$. For such strong $B$ electrons on Landau
levels feel weak Coulomb potential moving along the magnetic
field. In \cite{2} numerical solution of Schrodinger equation for
a hydrogen atom in strong $B$ was performed. According to this
solution the ground level goes to $-\infty$ when $B$ goes to
$+\infty$. However, the photon mass leads to the Coulomb potential
screening and the ground level remains finite at $B\to\infty$
\cite{3}. Since the electron at the ground Landau level moves
freely along the magnetic field, the problem resembles $D=2$ QED
and we will start our discussion from this theory.

\section{$D=2$ QED: screening of $\Phi$}

The following equation for an electric potential of the point-like
charge holds; see Fig. 1:

\begin{equation}
\Phi(\bar k) \equiv A_0(\bar k) = \frac{4\pi g}{\bar k^2} \; ;
\;\; {\bf\Phi} \equiv {\bf A}_0 = D_{00} + D_{00}\Pi_{00}D_{00} +
... \label{1}
\end{equation}

\bigskip

\begin{center}
\bigskip
\includegraphics[width=.8\textwidth]{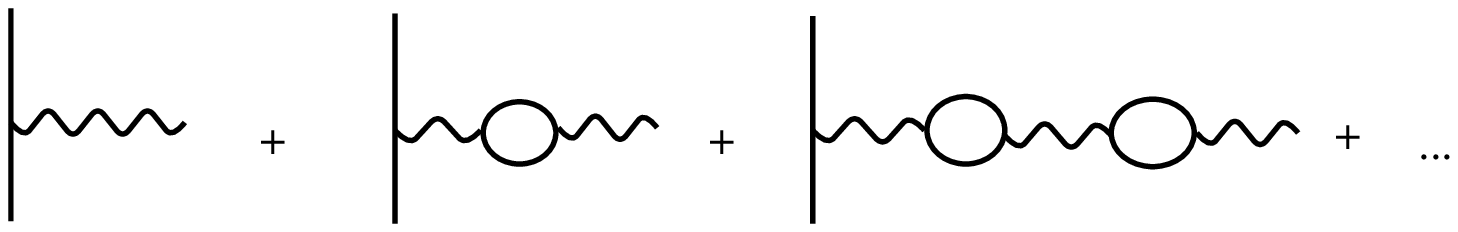}

Fig. 1. {\em Modification of the Coulomb potential due to the
dressing of the photon propagator.}

\end{center}

Summing the series we get:
\begin{equation} {\bf\Phi}(k) = -\frac{4\pi g}{k^2
+ \Pi(k^2)} \; , \;\; \Pi_{\mu\nu}
\equiv\left(g_{\mu\nu}-\frac{k_\mu k_\nu}{k^2}\right)\Pi(k^2) \;\;
, \label{2}
\end{equation}

\begin{equation}
\Pi(k^2) = 4g^2\left[\frac{1}{\sqrt{t(1+t)}}\ln(\sqrt{1+t} +\sqrt
t) -1\right] \equiv -4g^2 P(t) \;\; ,  \label{3}
\end{equation}

where $t\equiv -k^2/4m^2, \;\; [g] = $mass.

Taking $k = (0, k_\parallel)$, $k^2 = -k_\parallel ^2$ for the
Coulomb potential in the coordinate representation we get:
\begin{equation}
{\bf\Phi}(z) = 4\pi g \int\limits^\infty_{-\infty} \frac{e^{i
k_\parallel z} dk_\parallel/2\pi}{k_\parallel^2 + 4g^2
P(k_\parallel^2 /4m^2)} \;\; , \label{4}
\end{equation}
and the potential energy for the charges $+g$ and $-g$ is finally:
$ V(z) = -g{\bf\Phi}(z) \;\; . $

The asymptotics of $P(t)$ are:
\begin{equation}
P(t) = \left\{
\begin{array}{lcl}
\frac{2}{3} t & , & t\ll 1 \\
1 & , & t\gg 1 \;\; .
\end{array}
\right.\label{5}
\end{equation}

Let us take as an interpolating formula for $P(t)$ the following
expression:
\begin{equation}
\overline{P}(t) = \frac{2t}{3+2t} \;\; .\label{6}
\end{equation}
The accuracy of this approximation is not worse than 10\% for the
whole interval of $t$ variation, $0 < t < \infty$. Substituting an
interpolating formula in (\ref{4}) we get:
\begin{eqnarray}
{\bf\Phi} & = & 4\pi g\int\limits^{\infty}_{-\infty} \frac{e^{i
k_\parallel z} d k_\parallel/2\pi}{k_\parallel^2 +
4g^2(k_\parallel^2/2m^2)/(3+k_\parallel^2/2m^2)} = \nonumber
\\
& = & \frac{4\pi g}{1+ 2g^2/3m^2}
\int\limits_{-\infty}^{\infty}\left[\frac{1}{k_\parallel^2} +
\frac{2g^2/3m^2}{k_\parallel^2 + 6m^2 + 4g^2}\right]
e^{ik_\parallel z} \frac{dk_\parallel}{2\pi} = \\
&=& \frac{4\pi g}{1+ 2g^2/3m^2}\left[-\frac{1}{2}|z| +
\frac{g^2/3m^2}{\sqrt{6m^2 + 4g^2}} {\rm exp}(-\sqrt{6m^2
+4g^2}|z|)\right] \;\; . \nonumber \label{7}
\end{eqnarray}

In the case of heavy fermions ($m\gg g$)  the potential is given
by the tree level expression; the corrections are suppressed as
$g^2/m^2$.

In the case of light fermions ($m \ll g$):

\begin{equation}
{\bf\Phi}(z)\left|
\begin{array}{l}
~~  \\
m \ll g
\end{array}
\right. = \left\{
\begin{array}{lcl}
\pi e^{-2g|z|} & , & z \ll \frac{1}{g} \ln\left(\frac{g}{m}\right) \\
-2\pi g\left(\frac{3m^2}{2g^2}\right)|z| & , & z \gg \frac{1}{g}
\ln\left(\frac{g}{m}\right) \;\; .
\end{array}
\right. \label{8}
\end{equation}
$m=0$ corresponds to Schwinger model; photon gets mass.

Light fermions make transition from $m>g$ to $m=0$ continuous.

\section{$D=4$ QED}

In order to find the potential of a point-like charge we need an
expression for $P$ in strong $B$. One starts from electron
propagator $G$ in strong $B$. The solutions of Dirac equation in
the homogenious constant in time $B$ are known, so one can write
the spectral representation of the electron Green function. The
denominators contain $k^2-m^2-2neB$, and for $B>>m^2/e$ and
$k_\parallel^2<<eB$ in sum over levels the lowest Landau level
(LLL, $n=0$) dominates. In the coordinate representation a
transverse part of LLL wave function is: $\Psi\sim
exp((-x^2-y^2)eB)$ which in the momentum representation gives
$\Psi\sim exp((-k_x^2-k_y^2)/eB)$ (we suppose that $B$ is directed
along the $z$ axis).

Substituting the electron Green functions we get the expression
for the polarization operator in superstrong $B$.

For $B>>B_{cr}$, $k_\parallel^2 << eB$ the following expression is
valid \cite{4}:

\begin{eqnarray}
\Pi_{\mu\nu} & \sim & e^2 eB
\int\frac{dq_xdq_y}{eB}\exp(-\frac{q_x^2+q_y^2}{eB})* \nonumber \\
& * & \exp(-\frac{(q+k)_x^2+(q+k)_y^2}{eB})dq_0dq_z\gamma_{\mu}
\frac{1}{\hat q_{0,z}-m}(1-i\gamma_1 \gamma_2)\gamma_{\nu}*
\\
& * & \frac{1}{\hat q_{0,z}+\hat k_{0,z}-m}(1-i\gamma_1 \gamma_2)
= e^3 B * \exp(-\frac{k^2_\bot}{2eB}) *
\Pi_{\mu\nu}^{(2)}(k_\parallel\equiv k_z)\;\; . \nonumber
\label{9}
\end{eqnarray}

With the help of it the following result was obtained in \cite{1}:

\begin{equation}
{\bf\Phi}(k) =\frac{4\pi e}{k_\parallel^2 + k_\bot^2 + \frac{2 e^3
B}{\pi} {\rm exp}\left(-\frac{k_\bot^2}{2eB}\right)
P\left(\frac{k_\parallel^2}{4m^2}\right)} \;\; , \label{10}
\end{equation}
\begin{eqnarray}
{\bf\Phi}(z) & = & 4\pi e \int\frac{e^{ik_\parallel z} d
k_\parallel d^2 k_\bot/(2\pi)^3}{k_\parallel^2 + k_\bot^2 +
\frac{2 e^3B}{\pi} {\rm
exp}(-k_\bot^2/(2eB))(k_\parallel^2/2m_e^2)/(3+k_\parallel^2/2m_e^2)}
= \nonumber \\
& = & \frac{e}{|z|}\left[ 1-e^{-\sqrt{6m_e^2}|z|} +
e^{-\sqrt{(2/\pi) e^3 B + 6m_e^2}|z|}\right] \;\; . \label{11}
\end{eqnarray}

For the magnetic fields $B \ll 3\pi m^2/e^3$ the potential is
Coulomb up to small power suppressed terms:
\begin{equation}
{\bf\Phi}(z)\left| \begin{array}{l}
~~  \\
e^3 B \ll m_e^2
\end{array}
\right. = \frac{e}{|z|}\left[ 1+ O\left(\frac{e^3
B}{m_e^2}\right)\right] \label{12}
\end{equation}
in full accordance with the $D=2$ case, $e^3B \rightarrow g^2$.

In the opposite case of the superstrong magnetic fields $B\gg 3\pi
m_e^2/e^3$ we get:
\begin{equation}
{\bf\Phi}(z) = \left\{
\begin{array}{lll}
\frac{e}{|z|} e^{(-\sqrt{(2/\pi) e^3 B}|z|)} \; , \;\;
\frac{1}{\sqrt{(2/\pi) e^3 B}}\ln\left(\sqrt{\frac{e^3 B}{3\pi
m_e^2}}\right)>|z|>\frac{1}{\sqrt{e B}}\\
\frac{e}{|z|}(1- e^{(-\sqrt{6m_e^2}|z|)}) \; , \;\;  \frac{1}{m} >
|z|
> \frac{1}{\sqrt{(2/\pi)e^3 B}}\ln\left(\sqrt{\frac{e^3 B}{3\pi
m_e^2}}\right) \\
\frac{e}{|z|}\;\; , \;\;\;\;\;\;\;\;\;\;\;\;\;\;\;\;\;\;\;\;\;\;\;
  |z| > \frac{1}{m}
\end{array}
\right. \;\; , \label{13}
\end{equation}

\begin{equation}
 \bar V(z) = - e{\bf\Phi}(z) \;\; .
\label{14}
\end{equation}

In Fig. 2 the plot of a modified by the superstrong $B$ Coulomb
potential as well as its short- and long-distance asymptotics are
presented.

\bigskip

\begin{center}

\includegraphics[width=.6\textwidth]{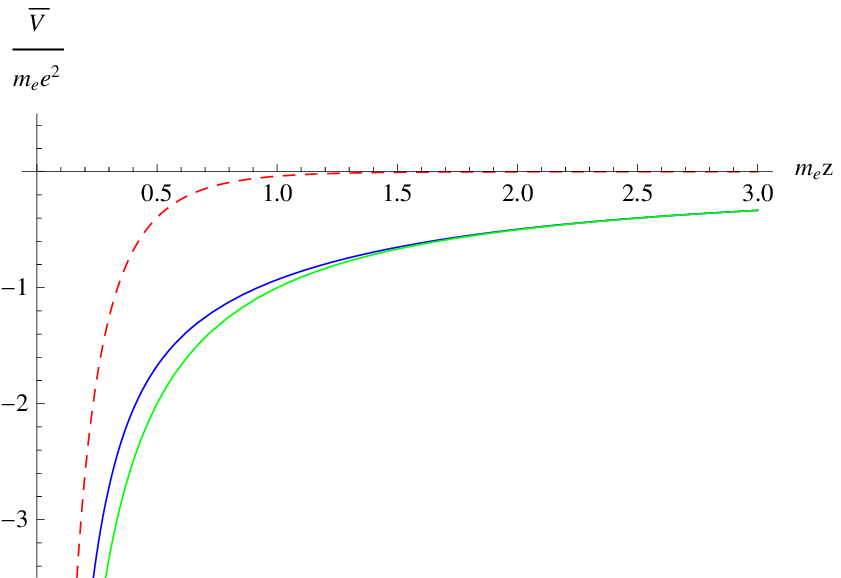}

Fig. 2. {\em A modified Coulomb potential at $B=10^{17}$G (blue,
dark solid) and its long distance (green, pale solid) and short
distance (red, dashed) asymptotics.}

\end{center}

\section{Electron in the magnetic field}

The spectrum of Dirac equation in the constant in time homogeneous
magnetic field is given by \cite{5}:
\begin{equation}
\varepsilon_n^2 = m_e^2 + p_z^2 + (2n + 1 + \sigma_z) \;{\rm  eB}
\;\; , \label{15}
\end{equation}
$n= 0, 1, 2, 3,...; \;\; \sigma_z = \pm 1$.

For $B > B_{cr} \equiv m_e^2/e$ the electrons are relativistic
with only one exception: the electrons from the lowest Landau
level (LLL, $n=0, \;\; \sigma_z = -1$) can be nonrelativistic.

In what follows we will find the spectrum of electrons from LLL in
the screened Coulomb field of the proton.

The spectrum of Schr\"{o}dinger equation in cylindrical
coordinates $(\rho, z)$ is \cite{6}:
\begin{equation}
E_{p_z n_\rho m \sigma_z} = \left(n_\rho + \frac{|m| +m+1+
\sigma_z}{2}\right)\frac{eB}{m_e} + \frac{p_z^2}{2m_e} \;\; ,
\label{16}
\end{equation}
LLL: $n_\rho = 0, \sigma_z = -1, m= 0, -1, -2,...$,
\begin{equation}
R_{0m}(\rho) = \left[\pi(2a_H^2)^{1+|m|} (|m|!)\right]^{-1/2}
\rho^{|m|}e^{(im\varphi - \rho^2/(4a_H^2))} \;\; . \label{17}
\end{equation}

Now we should take into account the electric potential of the
atomic nuclei situated at $\rho = z = 0$. For $a_H \ll a_B$
adiabatic approximation is applicable and the wave function in the
following form should be looked for:
\begin{equation}
\Psi_{n 0 m -1} = R_{0m}(\rho)\chi_n(z) \;\; , \label{18}
\end{equation}
where $\chi_n(z)$ is the solution of the Schr\"{o}dinger equation
for electron motion along the magnetic field:
\begin{equation}
\left[-\frac{1}{2m} \frac{d^2}{d z^2} + U_{eff}(z)\right]
\chi_n(z) = E_n \chi_n(z) \;\; . \label{19}
\end{equation}
Without screening the effective potential is given by the
following formula:
\begin{equation}
U_{eff} (z) = -e^2\int\frac{|R_{0m}(\rho)|^2}{\sqrt{\rho^2 +
z^2}}d^2 \rho \;\; , \label{20}
\end{equation}
 For $|z| \gg a_H$ the effective potential equals Coulomb:
\begin{equation}
U_{eff}(z) \left|
\begin{array}{l}
~~  \\
z \gg a_H
\end{array}
\right. = - \frac{e^2}{|z|} \;\; \label{21}
\end{equation}
and it is regular at $z=0$:
\begin{equation}
U_{eff}(0)
 \sim - \frac{e^2}{|a_H|} \;\; .
\label{22}
\end{equation}

Since $U_{eff}(z) = U_{eff}(-z)$, the wave functions are odd or
even under reflection $z\to -z$; the ground states (for $m=0, -1,
-2, ...$) are described by the even wave functions.

The energies of the odd states are:
\begin{equation}
E_{\rm odd} = -\frac{m_e e^4}{2n^2} + O\left(\frac{m_e^2
e^3}{B}\right) \; , \;\; n = 1,2, ... \;\; . \label{23}
\end{equation}
So, for the superstrong magnetic fields $B > m_e^2/e^3$ they
coincide with the Balmer series.

\section{Energies of even states; screening}

When screening is taken into account the expression for the
effective potential transforms into \cite{1}:
\begin{equation}
\tilde U_{eff} (z) = -e^2\int  \frac{|R_{0m}(\vec{\rho})|^2}
{\sqrt{\rho^2 +z^2}} d^2\rho \left[1-e^{-\sqrt{6m_e^2}\;z} +
e^{-\sqrt{(2/\pi)e^3 B + 6m_e^2}\;z}\right]
 \;\; . \label{24}
\end{equation}

For $m=0$ the following simplified formula can be used:
\begin {equation}
U_{simpl} (z) =  -e^2 \frac{1}{\sqrt{a_H^2 +z^2}}
\left[1-e^{-\sqrt{6m_e^2}\;z} + e^{-\sqrt{(2/\pi)e^3 B +
6m_e^2}\;z}\right]
 \;\; . \label{25}
\end{equation}

In Fig. 3 the plots of the effective potentials for $m=0$ and
$m=-1$ are presented.

\begin{center}

\includegraphics[width=.6\textwidth]{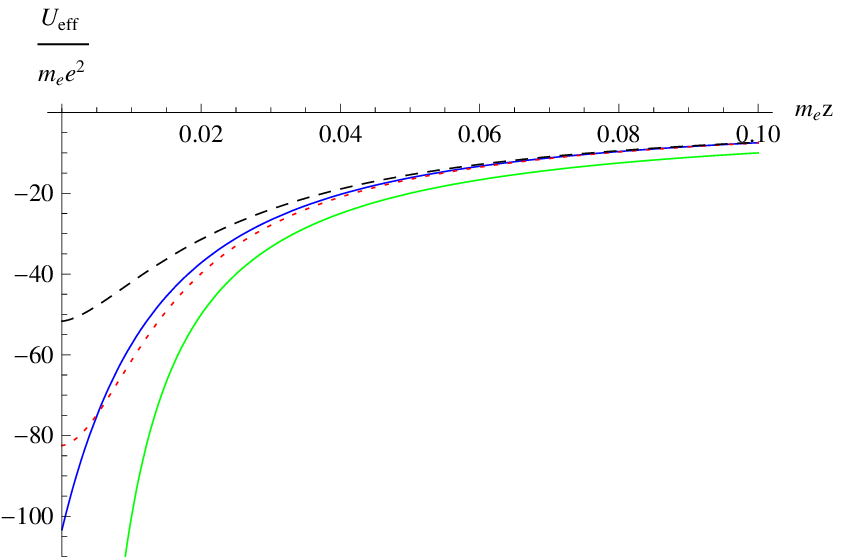}

Fig. 3. {\em Effective potential with screening for $m=0$ [dark
solid (blue) curve] and $m=-1$ (long-dashed curve), (\ref{24});
simplified potential  (short-dashed [red] curve) (\ref{25}). The
curves correspond to $B= 3 \times 10^{17}$G. Coulomb potential
(pale solid [green]) is also shown.}

\end{center}

\section{Karnakov - Popov equation}

It provides a several percent accuracy for the energies of even
states for  $H > 10^3$ ($H \equiv  B/(m_e^2 e^3$)), see \cite{7}.

The main idea is to integrate Shrodinger equation with the
effective potential from $x=0$ till $x=z$, where $a_H<<z<<a_B$ and
to equate the obtained expression for $\chi^\prime(z)$ to the
logarithmic derivative of Whittaker function - the solution of
Shrodinger equation with Coulomb potential, which exponentially
decreases at $z>>a_B$:
\begin{eqnarray}
&&2\ln\left(\frac{z}{a_H}\right) + \ln 2 - \psi(1+|m|) + O(a_H/z)
= \nonumber \\
&&  2\ln\left(\frac{z}{a_B}\right) + \lambda + 2\ln \lambda +
2\psi\left(1-\frac{1}{\lambda}\right) + 4\gamma + 2\ln 2 +
O(z/a_B) \;\; , \label{26}
\end{eqnarray}
where $\psi(x)$ is the logarithmic derivative of the
gamma-function and
\begin{equation}
E = -(m_e e^4/2)\lambda^2 \;\; . \label{27}
\end{equation}

\bigskip

\begin{center}

\includegraphics[width=.8\textwidth]{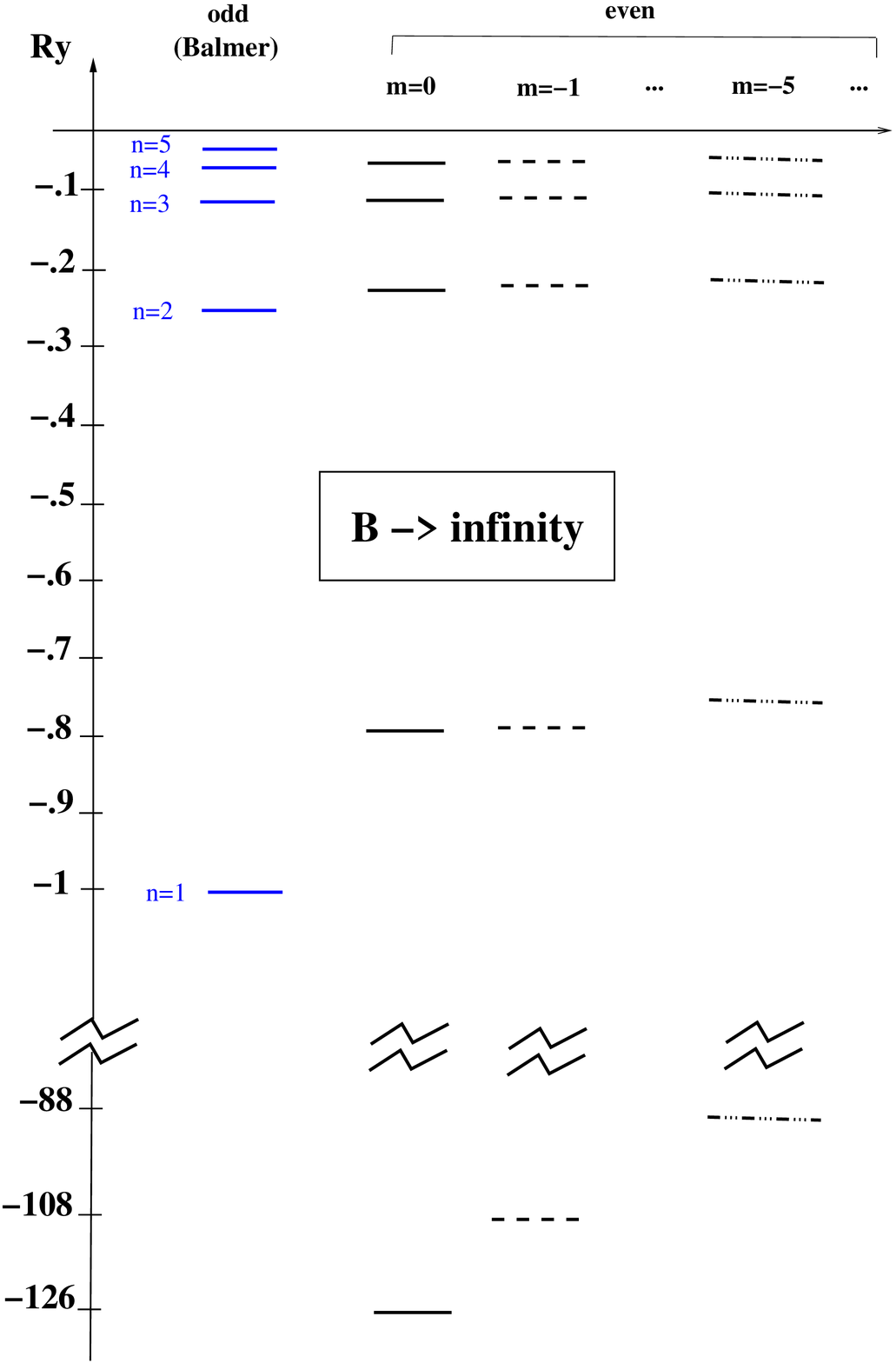}

Fig. 4. {\em Spectrum of the hydrogen levels in the limit of the
infinite magnetic field. Energies are given in Rydberg units, $Ry
\equiv 13.6 \; eV$.}

\end{center}

The modified KP equation, which takes screening into account,
looks like \cite{1}:
\begin{equation}
\ln\left(\frac{H}{1+\displaystyle\frac{e^6}{3\pi}H}\right) =
\lambda + 2\ln\lambda + 2\psi\left(1-\frac{1}{\lambda}\right) +
\ln 2 + 4\gamma + \psi(1+|m|) \;\; . \label{28}
\end{equation}

The spectrum of hydrogen atom in the limit $H\to\infty$ is shown
in Fig. 4.

\section{Conclusions}
\begin{itemize}
\item Atomic energies at superstrong $B$ is the only known (for
me) case when the radiative ``correction'' determines the energy
of states. \item The analytical expression  for the charged
particle electric potential in $d=1$ is given; for $m<g$ screening
takes place at all distances. \item The analytical expression for
the charged particle electric potential at superstrong $B$ in
$d=3$ is found; screening takes place at the distances
$|z|<1/m_e$. \item An algebraic formula for the energy levels of a
hydrogen atom originating from the lowest Landau level in
superstrong $B$ has been obtained.
\end{itemize}

\acknowledgments I am very grateful to the conference organizers
and to Mario Greco in particular for warm hospitality at La
Thuile. This work was partially supported by the grants RFFI
11-02-00441, N-Sh 4172.2010.2 and the contract 02.740.11.5158 of
the Ministry of Education and Science of the RF.


\begin{thebibliography}{0}
\bibitem{1} \BY{Vysotsky~ M.I.}
  \IN{JETP Lett.}{92}{2010}{15};
\BY{Machet~B. \atque Vysotsky~ M.I.} \IN{Phys. Rev.
D}{83}{2011}{025022}.
\bibitem{2} \BY{Elliott~R.J. \atque Loudon~R.} \IN{J. Phys. Chem.
Solids}{15}{1960}{196}.
\bibitem{3} \BY{Shabad~A.E. \atque Usov~V.V.} \IN{Phys. Rev.
Lett.}{98}{2009}{180403}; \IN{Phys. Rev. D}{77}{2008}{025001}.
\bibitem{4} \BY{Skobelev~V.V.} \IN{Izv. Vyssh. Uchebn. Zaved.,
Fiz.}{10}{1975}{142} [\IN{Sov. Phys. J.}{18}{1975}{1481}];
\BY{Loskutov~Yu.M. \atque Skobelev~V.V.} \IN{Phys.
Lett.}{56A}{1976}{151}.
\bibitem{5} \BY{Akhiezer~A.I. \atque Berestetskii~V.B.}
\TITLE{Quantum Electrodynamics}, in \TITLE{Interscience
publishers} (New-York -- London -- Sydney) 1965, p.~121.
\bibitem{6} \BY{Landau~L.D. \atque Lifshitz~E.M.} \TITLE{Quantum
Mechanics: Non-Relativistic Theory} (Pergamon, New York) 1991, 3rd
ed., Sec. 112, problem 1.
\bibitem{7} \BY{Karnakov~B.M. \atque Popov~V.S.} \IN{Zh. Eksp.
Teor. Fiz}{124}{2003}{996} [\IN{J. Exp. Theor.
Phys.}{97}{2003}{890}].
\end{thebibliography}
\end{document}